\documentstyle[12pt,a4wide,amssymb,epsfig]{article}
\begin{document}
\flushbottom
\renewcommand{\thefootnote}{\fnsymbol{footnote}}
\pagestyle{empty}
\setcounter{page}{0}
\begin{flushright}
DFTT 83/95
\\
hep-ph/9512353
\end{flushright}
\vspace*{1cm}
\begin{center}
\LARGE \bf
Neutrino oscillations
with three-generation
\\
\vspace{0.3cm}
mixings
and mass hierarchy\protect{\footnote{\normalsize Talk
presented by C. Giunti at TAUP 95,
Toledo (Spain), September 17-21, 1995}}
\vspace*{1cm}
\\
\Large \mediumseries
S.M. Bilenky$^{\mathrm{a}}$,
A. Bottino$^{\mathrm{b}}$,
C. Giunti$^{\mathrm{b}}$
and
C. W. Kim$^{\mathrm{c}}$
\\
\vspace{0.5cm}
\large \rm
\begin{tabular}{c}
$^{\mathrm{a}}$Joint Institute for Nuclear Research,
Dubna, Russia.
\\
$^{\mathrm{b}}$INFN and
Dipartimento di Fisica Teorica, Universit\`a di Torino,
\\
Via P. Giuria 1, 10125 Torino, Italy.
\\
$^{\mathrm{c}}$Department of Physics and Astronomy,
The Johns Hopkins University,
\\
Baltimore, Maryland 21218, USA.
\end{tabular}
\\
\vspace*{1cm}
Abstract
\\
\vspace{0.5cm}
\normalsize
\begin{minipage}[t]{14cm}
We have analyzed the results of
reactor and accelerator
neutrino oscillation experiments
in a model
with mixing of three neutrino fields
and
a neutrino mass hierarchy.
It is shown that
$ \nu_\mu \leftrightarrows \nu_e $
oscillations
with
$ 0.6 \le \Delta m^2 \le 100 \, \mathrm{eV}^2 $
and amplitude larger than
$ 2 \times 10^{-3} $
are not compatible with the existing limits
on neutrino oscillations
if the non-diagonal elements
of the mixing matrix are small.
Thus,
if the excess of positron events
recently observed in the LSND experiment
is due to
$ \bar\nu_\mu \to \bar\nu_e $
oscillations,
the mixing
in the lepton sector
is basically different from the CKM mixing
of quarks.
\end{minipage}
\end{center}

\newpage
\pagestyle{plain}
\setcounter{footnote}{0}
\renewcommand{\thefootnote}{\arabic{footnote}}

The LSND collaboration has recently reported
\cite{LSND}
to have found some positive indications
in favor of
$ \bar\nu_\mu \to \bar\nu_e $
transitions.
We will discuss here the implications
of this indication
in favor of neutrino oscillations
in the framework of a model
with mixing among three massive neutrino fields
and a mass hierarchy
that can allow to accommodate
the solar neutrino data
\cite{BBGK95}.

The probability of
$ \nu_{\alpha} \to \nu_{\beta} $
transitions
can be written in the following form:
$$
P_{\nu_\alpha\to\nu_\beta}
=
\left|
\sum_{k=2}^{3}
U_{\beta k}
\left(
{\mathrm{e}}^{ - i
{ \Delta m^2_{k1} L \over 2 p } }
- 1
\right)
U_{\alpha k}^{*}
+
\delta_{\beta\alpha}
\right|^2
$$
Here
$L$ is the distance between
the neutrino source and detector,
$ \Delta m^2_{k1} \equiv m^2_k - m^2_1 $
and
$p$ is the neutrino momentum.
We will assume that
$ m_1 \ll m_2 \ll m_3 $
and that
$  \Delta m^2_{21} $
is relevant for the possible suppression
of the solar $\nu_e$ flux.
In this case,
for experiments with terrestrial neutrinos
$ \Delta m^2_{21} L / 2 p \ll 1 $.
The probability
of
$ \nu_{\alpha} \to \nu_{\beta} $
transitions
and
the probability
of $ \nu_{\alpha} $
to survive
are given by
(see Ref.\cite{BFP92})
\arraycolsep=0cm
\begin{eqnarray*}
&&
P_{\nu_\alpha\to\nu_\beta}
=
{\displaystyle
A_{\nu_\alpha;\nu_\beta}
\over\displaystyle
2
}
\left(
1
-
\cos
{\displaystyle
\Delta m^2 L
\over\displaystyle
2 p
}
\right)
\;,
\hskip0.4cm
\alpha \not= \beta
\;,
%\label{E107}
\\
&&
P_{\nu_\alpha\to\nu_\alpha}
=
1
-
{\displaystyle
B_{\nu_\alpha;\nu_\alpha}
\over\displaystyle
2
}
\left(
1
-
\cos
{\displaystyle
\Delta m^2 L
\over\displaystyle
2 p
}
\right)
\;,
%\label{E109}
\end{eqnarray*}
with
$ \alpha , \beta = e , \mu , \tau $
and
$ \Delta m^2 \equiv \Delta m^2_{31} $.
Here
\arraycolsep=0cm
\begin{eqnarray}
&&
A_{\nu_\alpha;\nu_\beta}
=
4
\left| U_{\alpha3} \right|^2
\left| U_{\beta3} \right|^2
\label{E108}
\\
&&
B_{\nu_\alpha;\nu_\alpha}
=
\sum_{\beta\not=\alpha}
A_{\nu_\alpha;\nu_\beta}
=
4
\left| U_{\alpha3} \right|^2
\left(
1
-
\left| U_{\alpha3} \right|^2
\right)
\label{E110}
\end{eqnarray}
are the
amplitudes of oscillations.
Let us emphasize that
in the model under consideration
all the three oscillation channels
($\nu_{\mu}\leftrightarrows\nu_{e}$,
 $\nu_{\mu}\leftrightarrows\nu_{\tau}$,
 $\nu_{e}\leftrightarrows\nu_{\tau}$)
are open.
The oscillations in all channels
are characterized
by the {\em same} oscillation length
$
L_{\mathrm{osc}}
=
4 \pi p / \Delta m^2
$.
The oscillation amplitudes
(\ref{E108}) and (\ref{E110})
are determined by two mixing parameters,
which can be chosen to be
$ \left| U_{e3} \right|^2 $
and
$ \left| U_{\mu3} \right|^2 $.
{}From the unitarity of the mixing matrix
we have
$
\left| U_{\tau3} \right|^2
=
1
-
\left| U_{e3} \right|^2
-
\left| U_{\mu3} \right|^2
$.

In order to understand the implications
of the LSND positive signal
in the framework of this model,
we have analyzed
the negative results of all the other
neutrino oscillation experiments
with terrestrial neutrinos.
We have used the results of the
Bugey reactor experiment
\cite{BUGEY}
on the search for $\bar\nu_e$ disappearance,
of the
CDHS and CCFR accelerator experiments
\cite{CDHSCCFR}
on the search for $\nu_\mu$ disappearance,
of the
BNL E776 and KARMEN accelerator experiments
\cite{E776KARMEN}
on the search for
$ \nu_\mu \to \nu_e $
transitions
and of the
FNAL E531 accelerator experiment
\cite{E531}
on the search for
$ \nu_\mu \to \nu_\tau $
transitions.
For
$ \Delta m^2 $
we have considered the interval
$ 10^{-1} \, \mathrm{eV}^2 \le \Delta m^2 \le 10^{2} \, \mathrm{eV}^2 $,
that covers the LSND-allowed region
\cite{LSND}.
{}From the results of
reactor and accelerator disappearance experiments,
at fixed values of
$ \Delta m^2 $,
the allowed values of
the amplitudes
$ B_{\nu_e;\nu_e} $
and
$ B_{\nu_\mu;\nu_\mu} $
are constrained by
\begin{equation}
B_{\nu_e;\nu_e}
\le
B_{\nu_e;\nu_e}^{0}
\quad \mbox{and} \quad
B_{\nu_\mu;\nu_\mu}
\le
B_{\nu_\mu;\nu_\mu}^{0}
\;.
\label{E141}
\end{equation}

{}From Eqs.(\ref{E110}) and (\ref{E141})
it follows that the parameters
$ \left| U_{e3} \right|^2 $
and
$ \left| U_{\mu3} \right|^2 $
must be either small
or large (close to one).
For example,
at
$ \Delta m^2 = 6 \, \mathrm{eV}^2 $
we have
$
\left| U_{e3} \right|^2
\le
0.04
$
or
$
\left| U_{e3} \right|^2
\ge
0.96
$
and
$
\left| U_{\mu3} \right|^2
\le
0.02
$
or
$
\left| U_{\mu3} \right|^2
\ge
0.98
$.

Since
from the unitarity of the mixing matrix
the parameters
$ \left| U_{e3} \right|^2 $
and
$ \left| U_{\mu3} \right|^2 $
cannot be both close to one,
in the following we will consider
the three possible regions
for these parameters.

\noindent
{\bf I.}
$ \left| U_{e3} \right|^2 $
and
$ \left| U_{\mu3} \right|^2 $
are both small.

In this region
$ \left| U_{\tau3} \right|^2 $
is close to one
and from the results
of the FNAL E531 experiment
on the search for
$ \nu_\mu \to \nu_\tau $
transitions we obtain
rather strong limitations on the value of
$ \left| U_{\mu3} \right|^2 $
and,
consequently,
on the
$ \nu_\mu \to \nu_e $
oscillation amplitude
$ A_{\nu_\mu;\nu_e} $.
The results of our calculations
are presented in Fig.1
in the form of exclusion regions
in the
$ A_{\nu_\mu;\nu_e} $--$ \Delta m^2 $
plane.
The region allowed by the LSND experiment
is represented as the shadowed region limited by the two solid curves.
The curve passing through the filled circles
is the boundary of the excluded region
that was obtained from the results of the
Bugey,
CDHS
and
CCFR
disappearance experiments.
The curve passing through the open circles
was obtained by combining the results of the
Bugey experiment
and the FNAL E531
experiment
on the search for
$ \nu_\mu \to \nu_\tau $
transitions.
The dashed line reproduces the results of the
Bugey disappearance experiment
(we took into account the fact that
$ A_{\nu_\mu;\nu_e} \le B_{\nu_e;\nu_e} $).
The dotted and dash-dotted lines
reproduce,
respectively,
the results of the
BNL E776
and
KARMEN
experiments
on the search for
$ \nu_\mu \to \nu_e $
transitions.

{}From Fig.1
it can be seen that,
in the framework of the model under consideration,
the results of
$\nu_e$ and $\nu_\mu$
disappearance experiments
together with
the results of the FNAL E531
$\nu_\mu\to\nu_\tau$
appearance experiment
provide more severe restrictions
on
$ A_{\nu_\mu;\nu_e} $
than the results
of direct
$\nu_\mu\to\nu_e$
appearance experiments.
Furthermore,
practically all the LSND-allowed
region
is inside the region which is forbidden by
the results of the other reactor and accelerator experiments.
Therefore,
if the results of the LSND experiment
will be confirmed,
it will mean that
the parameters
$ \left| U_{e3} \right|^2 $
and
$ \left| U_{\mu3} \right|^2 $
cannot be both small.

\noindent
{\bf II.}
$ \left| U_{e3} \right|^2 $
is large
and
$ \left| U_{\mu3} \right|^2 $
is small.

This region is forbidden by the solar neutrino data.
In fact,
in the case of a neutrino mass hierarchy,
the survival probability of the solar neutrinos
is given by
\cite{SOLARTHREEGEN}
$$
P_{\nu_e\to\nu_e}
=
\left(
1
-
\left| U_{e3} \right|^2
\right)^2
P_{\nu_e\to\nu_e}^{(1,2)}
+
\left| U_{e3} \right|^4
\;,
$$
where
$ P_{\nu_e\to\nu_e}^{(1,2)} $
is the survival probability
due to the mixing between
the first and the second generations.
In the interval of $ \Delta m^2 $
under consideration,
we have
$
P_{\nu_e\to\nu_e}
\ge
0.92
$
for all values of the neutrino energy.
With such a high lower bound for the $\nu_e$ survival probability
it is not possible to explain
the results of the
solar neutrino experiments
\cite{SOLAREXP}.

\noindent
{\bf III.}
$ \left| U_{e3} \right|^2 $
is small
and
$ \left| U_{\mu3} \right|^2 $
is large.

In this region,
in the linear approximation
over the small parameters
$ \left| U_{e3} \right|^2 $
and
$ ( 1 - \left| U_{\mu3} \right|^2 ) $
we have
\arraycolsep=0cm
\begin{eqnarray}
&&
A_{\nu_\mu;\nu_e}
\simeq
4
\left| U_{e3} \right|^2
\;,
\label{E185}
\\
&&
A_{\nu_\mu;\nu_\tau}
\simeq
4
\left(
1
-
\left| U_{\mu3} \right|^2
-
\left| U_{e3} \right|^2
\right)
\;,
\label{E186}
\\
&&
A_{\nu_e;\nu_\tau}
\simeq
0
\;.
\label{E190}
\end{eqnarray}
{}From Eq.(\ref{E190})
and the unitarity relations (\ref{E110})
it follows that
\arraycolsep=0cm
\begin{eqnarray}
&&
A_{\nu_\mu;\nu_e}
\simeq
B_{\nu_e;\nu_e}
\;,
\label{E187}
\\
&&
A_{\nu_\mu;\nu_\tau}
\simeq
B_{\nu_\mu;\nu_\mu}
-
B_{\nu_e;\nu_e}
\;.
\label{E188}
\end{eqnarray}

In Ref.\cite{BBGK95}
we have shown that
in region III
the LSND positive signal
is compatible with the negative results
of all the other experiments
on the search for neutrino oscillations.
Thus,
a confirmation of the LSND result will
constrain the parameters
$ \left| U_{e3} \right|^2 $
and
$ \left| U_{\mu3} \right|^2 $
to lie in region III.
In this case
$ \nu_\mu \to \nu_\tau $
oscillations
that are searched for
in the CHORUS and NOMAD experiments
\cite{CHORUSNOMAD}
are allowed
and
$ \nu_e \to \nu_\tau $
oscillations
are strongly suppressed
(for example,
$ A_{\nu_e;\nu_\tau} \lesssim 3 \times 10^{-5} $
at
$ \Delta m^2 = 6 \, \mathrm{eV}^2 $).
If a positive signal
will be found in both
$ \nu_\mu \to \nu_e $
and
$ \nu_\mu \to \nu_\tau $
channels,
using
Eqs.(\ref{E185}) and (\ref{E186})
it will be possible to determine
the values of the mixing parameters
$ \left| U_{\mu3} \right|^2 $
and
$ \left| U_{e3} \right|^2 $.
In this case,
the validity of the model under discussion
can be tested by checking
the presence of only one relevant mass scale
$ \Delta m^2 $
for all the oscillation channels.

In conclusion,
we have considered
a model
with mixing of three neutrino fields
and
a neutrino mass hierarchy that
can accommodate the results of
the solar neutrino experiments.
After the calibration of the GALLEX detector
with a radioactive source
\cite{GALLEX}
the indications in favor of neutrino oscillations
coming from solar neutrino experiments
have become more significant.
We think that
the model that we have considered is
the simplest and most realistic
model of neutrino mixing.
It seems very appropriate to
analyze in the framework of this model
all the data from the existing experiments
on the search for neutrino oscillations
and
to infer predictions for the results
of future experiments
(see Ref.\cite{LISI}).
Our analysis has shown that
a confirmation of the LSND positive result
for
$ \nu_\mu \to \nu_e $
oscillations
would imply that
$ \left| U_{e3} \right|^2 $
is small
and
$ \left| U_{\mu3} \right|^2 $
is large
in the scheme under discussion.
This result would mean that
neutrino mixing is basically different
from quark mixing:
there is no natural hierarchy of coupling
in the lepton sector
and $\nu_\mu$ is the
``heaviest'' neutrino.

\begin{figure}[p]
\mbox{\epsfig{file=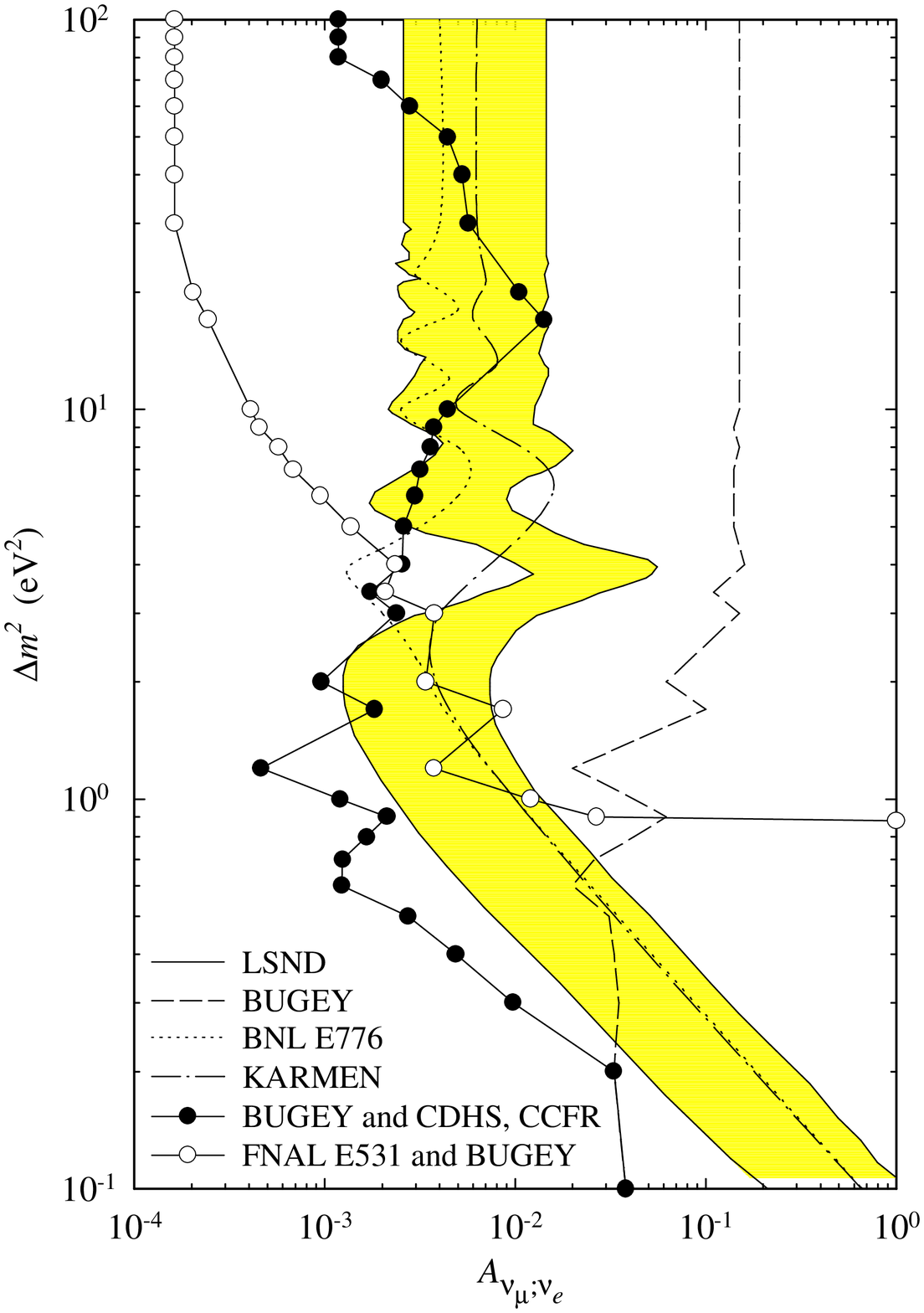,height=0.9\textheight}}
\centerline{Figure 1}
\label{FIG1}
\end{figure}


\begin{thebibliography}{99}

\bibitem{LSND}
C. Athanassopoulos et al.,
Phys. Rev. Lett. 75 (1995) 2650.

\bibitem{BBGK95}
S.M. Bilenky et al.,
Phys. Lett. B 356 (1995) 273.

\bibitem{BFP92} \label{BFP92}
A. De Rujula et al.,
Nucl. Phys. B 168 (1980) 54;
V. Barger and K. Whisnant,
Phys. Lett. B 209 (1988) 365;
S.M. Bilenky et al.,
{\it ibid} 276 (1992) 223.

\bibitem{BUGEY}
B. Achkar et al.,
Nucl. Phys. B 434 (1995) 503.

\bibitem{CDHSCCFR}
F. Dydak et al.,
Phys. Lett. B 134 (1984) 281;
I.E. Stockdale et al.,
Phys. Rev. Lett. 52 (1984) 1384.

\bibitem{E776KARMEN}
L. Borodovsky et al.,
Phys. Rev. Lett. 68 (1992) 274;
B. Armbruster et al.,
Nucl. Phys. B (Proc. Suppl.) 38 (1995) 235.

\bibitem{E531}
N. Ushida
Phys. Rev. Lett. 57 (1986) 2897.

\bibitem{SOLAREXP}
B.T. Cleveland et al.,
Nucl. Phys. B (Proc. Suppl.) 38 (1995) 47;
K. S. Hirata et al.,
Phys. Rev. D 44 (1991) 2241;
GALLEX Coll.,
Phys. Lett. B 357 (1995) 237;
V.N. Gavrin,
Talk presented at TAUP 95
Toledo (Spain), Sept. 1995.

\bibitem{SOLARTHREEGEN}
T.K. Kuo and J. Pantaleone,
Phys. Rev. Lett. 57 (1986) 1805;
X. Shi and D.N. Schramm,
Phys. Lett. B 283 (1992) 305.

\bibitem{CHORUSNOMAD}
K. Winter,
Nucl. Phys. B (Proc. Suppl.) 38 (1995) 211;
L. Di Lella,
Nucl. Phys. B (Proc. Suppl.) 31 (1993) 319.

\bibitem{GALLEX}
GALLEX Coll.,
Phys. Lett. B 342 (1995) 440.

\bibitem{LISI}
G.L. Fogli, E. Lisi and D. Montanino,
Phys. Rev. D 49 (1994) 3626;
CERN-TH-7491-94.

\end{thebibliography}
\end{document}